# Structured Light Modal Interface via Liquid-Crystal Planar Optics

Chun-Yu Li[1†], Si-Jia Liu[2†], Hai-Jun Wu[1], Jia-Qi Jiang[1], Bo Zhao[1], Carmelo Rosales-Guzmán[1,3], Zhi-Han Zhu[1*], Peng Chen[2*], and Yan-Qing Lu[2*]

[1] Wang Da-Heng Center, Harbin University of Science and Technology, Harbin 150080, China
[2] National Laboratory of Solid State Microstructures, Key Laboratory of Intelligent Optical Sensing and Manipulation, College of Engineering and Applied Sciences, and Collaborative Innovation Center of Advanced Microstructures, Nanjing University, Nanjing 210093, China.
[3] Centro de Investigaciones en Óptica, A.C., Loma del Bosque 115, Colonia Lomas del Campestre, 37150 León, Gto., Mexico
* e-mail: zhuzhihan@hrbust.edu.cn, chenpeng@nju.edu.cn, and yqlu@nju.edu.cn
† These authors contributed equally.

**Abstract**: Recent advances in planar optics with geometric-phase superstructures have brought a new paradigm in the control of structured light and, in particular, has substantially enhanced the capabilities of generating and detecting orbital angular momentum (OAM) states of light and associated spatial modes. However, the structured modal interface that can reciprocally link OAM states via adiabatic control and access-associated higher-order geometric phase remains absent in planar optics. In this work, we propose and experimentally demonstrate a planar optical astigmatic retarder fabricated with liquid-crystal (LC) geometric phase. The LC superstructure was designed with the principle of fractional Fourier transformation and is capable of reciprocal conversion between all possible OAM states on the same modal sphere. Such a planar device paves the way towards an easily deployed modal interface of paraxial OAM states, unlocks the resource of higher-order geometric phase, and has promising applications in high-dimensional classical/quantum information.

## Introduction

In quantum theory, complex vectors termed quantum states or wavefunctions are used to describe the existence and evolution of entities within corresponding degrees of freedom [1-3]. Devices that perform unitary transformation (or evolution) for such vectors, i.e., pointing control without changing vector length in the Hilbert space, are broadly called retarders. The essence of this adiabatic control lies in introducing phase retardance between orthogonal basis in the Hilbert space, which also gives rise to the origin of geometric phase [4-6]. For instance, the best-known half- and quarter-wave plates used to control the polarization states of light fields are birefringence retarders providing $\lambda/2$ and $\lambda/4$ phase shifts, respectively [7]. Beyond a binary system, the adiabatic control of states in a high-dimensional space offers a richer source to explore new physics and applications [8,9]. Taking recent optics as an example, high-dimensional photonic states built by spatially structured light and orbital angular momentum (OAM) degrees of freedom have brought fresh new ideas to many areas of research, such as quantum/classical communication, optical tweezers, and superresolution imaging [10-17]. Notably, most advancements were made by exploiting larger encoding spaces or unique light-matter interfaces enabled by OAM states and associated structured light, but few directly benefited from high-dimensional unitary transformations and associated higher-order geometric phases. One important reason is that astigmatic retarders used for the unitary transformation of OAM states carried by spatial modes lag the current requirements of experimental studies [18,19]. Astigmatic retarders widely used in present studies suffer from complex lens systems, leading to low accuracy, which in addition are difficult to manipulate in experiments, despite the efforts to reduce the size of the setup [20]. In comparison, great progress has been made in the generation and detection of OAM states and associated spatial modes with planar optics [21-23]. These popular planar elements, carrying customized geometric phase patterns, are commonly fabricated with microstructured liquid crystals (LCs) or dielectric metasurfaces [24-27]. To bridge this gap, we present an astigmatic retarder based on planar optics with an LC geometric phase, whose microstructure (or geometric-phase pattern) was designed to provide a phase shift with exactly one-quarter of a wave between orthogonal Hermite-Gaussian modes. The retarder can work as a true zero-order '$\lambda/4$ waveplate' for OAM state control, which can freely link states on the same modal sphere formed by all possible Hermite-Laguerre-Gaussian (HLG) modes. To demonstrate this principle, we experimentally show its good performance in unitary transformation of the OAM state and associated spatial modes. The LC geometric phase was realized with nematic LCs via the photoalignment technique, which promises high precision, high efficiency, and high reliability of the astigmatic retarder [28]. Furthermore, we analyze the OAM conservation of the astigmatic transformation system and observe the higher-order geometric phase shift resulting from cyclic transformation on the modal sphere.

## Concept & Principle

The two most common families of structured Gaussian beams — Laguerre-Gauss and Hermite-Gauss modes (denoted as $LG_{\ell,p}$ and $HG_{m,n}$) — are eigen solutions of the paraxial wave equations in cylindrical and Cartesian coordinates, respectively [29]. Complex vector space regarding OAM states formed by these paraxial spatial modes have an SU(2) modal structure and geometric representation similar to those of polarization states [30]. In particular, as shown in Fig. 1(a), an analog of the Poincaré sphere is widely used to represent geometrically the modal structure of OAM states formed by first-order ($N=1$) Hermite-Laguerre-Gauss (HLG) modes, where circular polarizations (carrying spin angular momenta $\pm 1\hbar$) on the two poles are replaced by conjugate LG modes with opposite topological charges $\ell = \pm 1$. Further, the mutually unbiased bases (MUBs) on the equator, i.e., rotated linear polarizations, are replaced by rotated $HG_{1,0}$ modes [31]. Notably, this fairly consistent analogy is not valid for higher-order cases with $N = 2p + |\ell| = m + n > 1$. Taking three 4$^\text{th}$-order modal spheres as examples, as shown in Fig. 1(b), the higher-order HG modes on the equator are no longer MUBs of LG modes on the poles but are superpositions of all possible LG modes with order $N = 4$, including those that are not on the sphere [30]. It is even more unusual for the particular case $\ell = 0$. Due to the lack of the helicity degree of freedom, all possible relevant modes can only form a half modal sphere. Another point to be aware of is that the amount of geometric phase resulting from a cyclic transformation on a modal sphere depends only on the parameter $|\ell|$ of the LG modes on the two poles, i.e., not on the order $N$, which will be experimentally examined later.

The physical foundation of the unitary transformation of OAM states on the modal sphere derives from the underlying SU(2) structure of HLG modes, which is also true for the recently revealed multi-path HLG modes [32-34]. For an easier understanding, we interpret the mechanism via a crucial mathematical relation arising from Hermite and Laguerre polynomials, i.e., a diagonally placed $HG_{m,n}^{45°}$ and $LG_{\pm\ell,p}$ on the same modal sphere have the same modal spectral density when decomposed with all $N$-order HG modes [18,35,36]. The only difference appears in the relative phase between successive HG components, making the adiabatic control of states on the sphere possible. A detailed formulism description is provided in the Supplementary Material (SM). Here, we show the relation with the simplest case on the 1$^\text{st}$-order modal sphere in Fig. 1(a), where the two decompositions can be expressed as

$$\text{HG}_{1,0}^{45°} = \sqrt{\frac{1}{2}}(\text{HG}_{1,0} + \text{HG}_{0,1}), \quad (1)$$

$$\text{LG}_{1,0} = \sqrt{\frac{1}{2}}(\text{HG}_{1,0} + i\text{HG}_{0,1}). \quad (2)$$

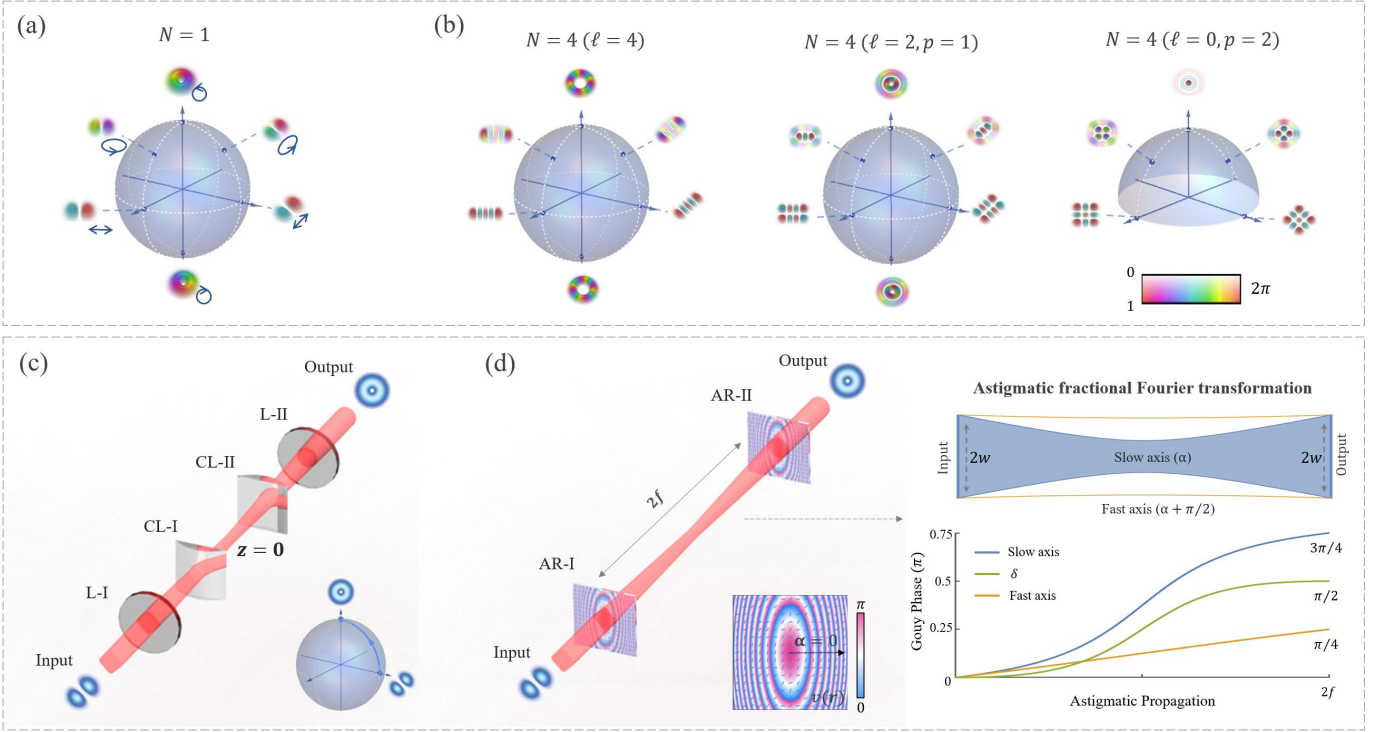

**Figure 1.** Concept and principle. (a) Poincaré sphere and its HLG analog with $N = 1$, on which blue vector ellipses and rainbow patterns denote polarizations and corresponding HLG modes, respectively; (b) three 4th-order modal spheres, where a 2D color map shown in the right bottom is used to visualize the complex amplitude (including intensity and wavefront) patterns on the spheres; (c) principle of astigmatic modal interface via the lens approach, where L and CL are the lens and cylindrical lens, respectively; and (d) modal interface via the planar optics approach, where the LC astigmatic retarder (AR) was designed with the principle of fractional Fourier transformation (right side).

The result indicates that the reciprocal conversion $\text{HG}_{m,n}^{45°} \leftrightarrow \text{LG}_{\ell,p}$ (where $\ell = m - n$) requires introducing a $\pi/2$ phase shift between the successive HG components. The Gouy phase provides a natural approach for this adiabatic control. More specifically, upon propagation the Gouy phase of HG modes increases with the order $N$ and, more importantly, it is axially separable in Cartesian coordinates, which can be expressed as follows:

$$(N+1)\phi = \left(m + \frac{1}{2}\right)\phi_x + \left(n + \frac{1}{2}\right)\phi_y, \quad (3)$$

where $\phi$ and $\phi_{x,y}$ denote the Gouy phase and its axial components, respectively. Eq. (3) indicates that we can use a pair of cylindrical lenses to introduce the $\pi/2$ phase retardance, as shown in Fig. 1(c). Because this phase retardance is offered by astigmatic imaging, the conversion is thus broadly called astigmatic transformation [18,19]. Notably, the center of the cylindrical lens pair shown in Fig. 1(c) must be placed at the beam waist plane ($z = 0$). For this reason, in practice, it is inevitable to use another pair of relay lenses that cooperates with astigmatic imaging, making the configuration of the whole retarder more complicated.

The planar astigmatic retarder in this work was designed with the principle of fractional Fourier transformation [37], as shown in Fig. 1(d). The LC director orientation distribution $v(\boldsymbol{r})$ enables a point-by-point control of polarization along the transverse plane and leads to a spin-dependent wavefront control, i.e., spatial light modulation enabled by the geometric phase of microstructured birefringence. The spatial light modulation follows the relation $\hat{e}_\pm \to \exp[\pm 2iv(\boldsymbol{r})]\hat{e}_\mp$ as global birefringence working at the half-wave condition, where $\hat{e}_\pm$ denotes left/right circular polarization [28]. The term 'fractional Fourier transformation' refers to the design of the device, which is capable to induce independent fractional Gouy phases along $x$ and $y$ directions to an input light field, where $\phi_y = \pi/4$ and $\phi_x = 3\pi/4$ (more details can be found in SM). Thus, the astigmatic phase shift created (i.e., $\delta = \phi_x - \phi_y$) is exactly one-quarter of a wave, enabling the reciprocal conversion $\text{HG}_{m,n}^{45°}\hat{e}_+ \leftrightarrow \text{LG}_{\ell,p}\hat{e}_-$ for collimated Gaussian beams, in a similar way to a zero-order $\lambda/4$ waveplate used for polarization control. Compared to the lens approach shown in Fig. 1(c), this LC geometric phase empowered astigmatic transformation only consists of two identical planar optical elements. Thus, the system exhibits easy alignment and is insensitive to off-axis errors.

**Results & Discussion**

**Experimental Setup** — To prove this principle, we first demonstrate the performance of the planar astigmatic retarder in the unitary transformation of HLG modes. Figure 2(a) shows the schematic setup of the apparatus, which adopts a configuration of a polarization Michelson interferometer capable of astigmatic conversion for both scalar and vectorial beams. To be more specific, assuming the input is a vectorial mode composed of orthogonal HG modes, e.g., $1/\sqrt{2}\left(\text{HG}_{m,n}\hat{e}_+ + \text{HG}_{n,m}\hat{e}_-\right)$, its two polarization components would be first separated along the two arms of the interferometer by the polarizing beam splitter (PBS) for independent astigmatic conversion, or only one arm would work in the presence of a scalar beam with $\hat{e}_+$ or $\hat{e}_-$ polarization. Each arm contains an LC astigmatic retarder combined with a Faraday rotator and quarter-wave plates, the distance between the LC element and the mirror is set as exactly one slow-axis focal length of the retarder. Figure 2(b) shows the experimental characterizations of the LC fabricated device, including a polarizing micrograph and the measured LC director distribution via spatial Stokes polarimetry [38]. The space-variant brightness in the polarized micrograph indicates the desired variation of LC directors and the measured LC orientation distribution is highly consistent with our design. Under this configuration, the HG mode in each arm would be converted into an HLG mode depending on the relative angle between the slow-axis angle of the retarder ($\alpha$) and the major axis of HG modes ($\beta$), i.e., $\text{HG}_{m,n}^{(\beta)}\hat{e}_\pm \to \text{HLG}_{m,n}^{(\alpha-\beta)}\hat{e}_\mp$. An example of a unitary transformation path (rainbow arrow) on the modal sphere versus $\alpha$ is given in the bottom right of Fig. 2(a). In particular, as $\alpha - \beta = 45°$, the output would become a cylindrical vector mode with the state $1/\sqrt{2}\left(\text{LG}_{+\ell,p}\hat{e}_+ + \text{LG}_{-\ell,p}\hat{e}_-\right)$.

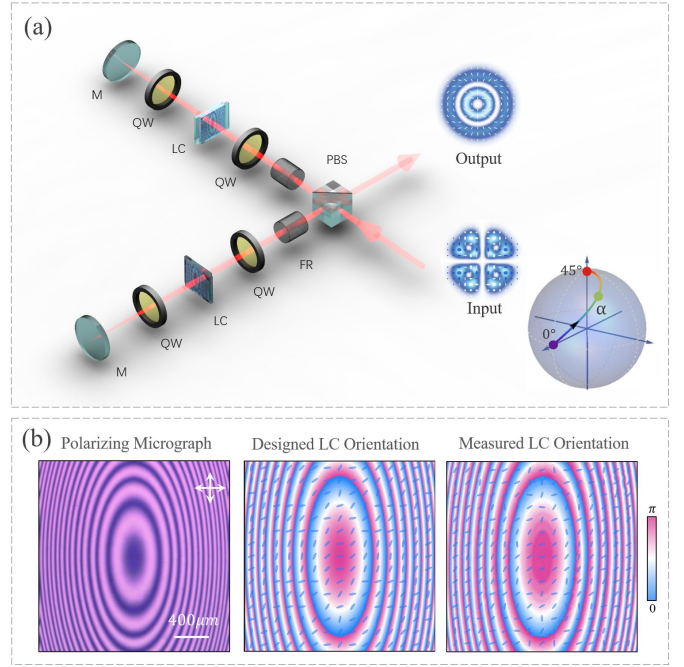

**Figure 2.** (a) Experimental setup of the vectorial modal interface, where key components include a polarizing beam splitter (PBS), Faraday rotator (FR), quarter-wave plates (QWP), mirrors (M) and LC retarders; and (b) LC device characterizations including a polarizing micrograph taken at 0 V under crossed polarizers and the measured LC orientation via spatial polarimetry.

In experiments, we first examined the astigmatic transformation of scalar HLG modes, in which several $\text{HG}_{m,n}^{0°}$ modes with orders $N = m + n$ from 0 to 4 were prepared as input signals to be converted, as shown in the complex amplitude patterns in the first row of Fig. 3(a). The second and third rows show their conversions corresponding to $\alpha = 22.5°$ and $45°$, respectively. All observed patterns agree well with their theoretical reference provided in SM, which confirms the performance and precision of the planar astigmatic retarder. Additional data involving astigmatic transformation of all possible $\text{HG}_{m,n}^{0°}$ modes with $N = m + n \leq 4$ are provided in SM. On this basis, we further consider additional groups of vector modes that can be represented geometrically as four points on the spin-orbit hybrid modal sphere spanned by $\text{HG}_{3,1}\hat{e}_+$ and $\text{HG}_{1,3}\hat{e}_-$, as shown in Fig. 3(b). The results show that these four states were converted to states on a new hybrid sphere spanned by $\text{HLG}_{3,1}^\alpha \hat{e}_+$ and $\text{HLG}_{1,3}^\alpha \hat{e}_-$ and remained at their original positions. In particular, as $\alpha = 45°$, the output states become the more common cylindrical modes on the hybrid sphere with OAM $\ell = 2$ defined by $\text{LG}_{+2,1}\hat{e}_+$ and $\text{LG}_{-2,1}\hat{e}_-$.

**OAM Conservation** — When we reexamine the above results from a deeper physical insight, beyond the technical aspect, interesting issues arise: *what mechanism causes this high-dimensional unitary transformation within spatial degrees of freedom, leading to both spatial amplitude and phase that are drastically changed via an adiabatic evolution, and how does the OAM light–matter exchange occur that can maintain the system OAM conservation?* The issues can be elucidated by investigating the modal evolution of a beam during astigmatic operation. We will see that, i) the OAM exchange occurs only in the LC retarder at the HG port and ii) the mechanism here is extremely similar to that of the so-called 'time lens' used for pulse duration transformation [39,40].

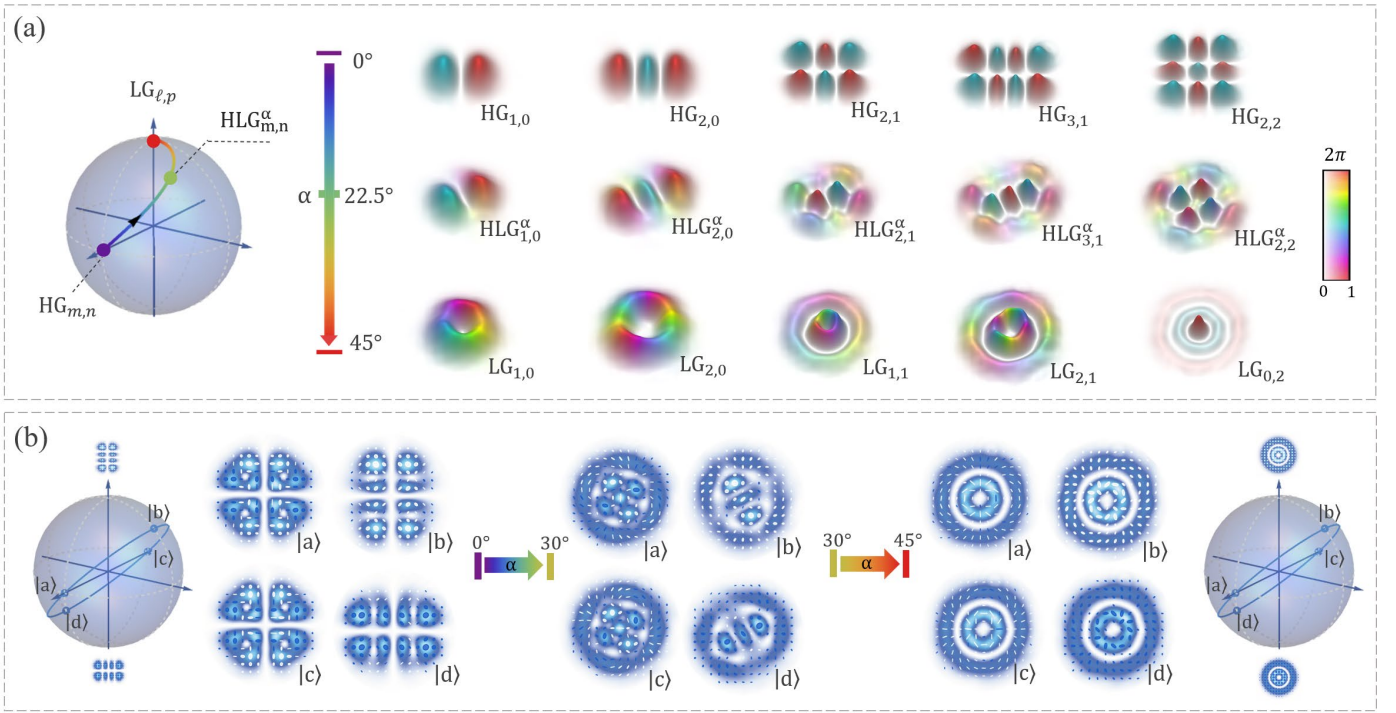

**Figure 3.** Experimental results of scalar (a) and vector (b) modal conversions, where the rainbow arrow denotes the LC-retarder angle ($\alpha$) and associated positions of converted states on the modal sphere. The spatial complex amplitude and vectorial profiles are observed using the technique in ref. 38.

Taking the conversions in Fig. 4(a) as an example, the first LC retarder provides an astigmatic wavefront $2v(\boldsymbol{r},\alpha)$ on the input $\mathrm{HG}_{m,n}^{\beta}$ mode (here, assuming $\beta=0°$ for simplicity) and makes it no longer a paraxial eigenmode. As a result, this astigmatic beam would continuously change in its spatial complex amplitude upon propagation and finally evolve into the desired $\mathrm{HLG}_{m,n}^{(\alpha-\beta)}$ at the plane $z=2f$ but still carry a residual astigmatism $2v^*(\boldsymbol{r},\alpha)$ that will be removed by the second LC retarder. On this basis, a quantitative understanding is achieved as we revisit the whole diffraction from the perspective of OAM spectrum evolution, i.e., $\sum c_{\ell,p}\mathrm{LG}_{\ell,p}$ dynamics, see SM for details. The first astigmatic operation stretches the OAM spectrum of the passing beam, and more importantly, the resulting asymmetric spectrum about the axis $\ell=0$ indicates that the passing beam $\mathrm{HG}_{m,n}^{\beta}(\boldsymbol{r})e^{i2v(\boldsymbol{r},\alpha)}$ has already carried net OAM with an average amount of $\bar{\ell}\hbar$ per photon given by

$$\bar{\ell} = |m-n|\sin[2(\alpha-\beta)]. \quad (4)$$

Although the power spectrum $c_{\ell,p}^2$ is propagation-invariant, the asynchronous Gouy phase continuously modulates the intramodal phase between successive LG components [41-44]. This Gouy-phase-mediated diffraction reshapes the beam structure into an exact $\mathrm{HLG}_{m,n}^{(\alpha-\beta)}e^{i2v^*(\boldsymbol{r},\alpha)}$ before the second astigmatic operation. Afterwards, the stretched OAM spectrum is compressed by removing its residual astigmatism after passing the second LC retarder at the plane $z=2f$. In particular, as $\alpha-\beta=45°$, the spectrum is compressed into a single value $\ell=m-n$, and the output is exactly an $\mathrm{LG}_{\ell,p}$ mode. In general, this two-stage adiabatic operation of the spatial spectrum in terms of the OAM can be regarded as a two-dimensional analogous of the so-called time lens technique, which is used to adiabatically control the temporal spectrum of light [39,40].

**Higher-Order Geometric Phase** — The high-dimensional unitary transformation demonstrated above paves the way towards the exploitation of higher-order geometric phases in terms of the OAM degree of freedom. For this, we also experimentally measured the higher-order geometric phase with our LC planar device. The amount of phase shift depends on both the topological charge of polar LG conjugates and the transformation path on the sphere, given by

$$\phi_{gp} = (|\ell|+1)\Omega/2 \quad (5)$$

where $\Omega$ is the solid angle encompassed by the geodesic path drawn by the cyclic transformation. Here, we adopt a self-reference method to conveniently observe the phase shift resulting from cyclic transformations. The experimental details are provided in SM. Specifically, LG modes as signals were first combined with a reference $\mathrm{LG}_{00}$ mode via orthogonal polarizations, forming vector modes expressed as $1/\sqrt{2}\left(\mathrm{LG}_{+\ell,p}\hat{e}_+ + \mathrm{LG}_{0,0}\hat{e}_-\right)$. After a cyclic transformation with a solid angle $\Omega$ (realized with the $\lambda/4$ - $\lambda/2$ - $\lambda/4$ combination control), these vector modes evolve into $1/\sqrt{2}\left(\mathrm{LG}_{+\ell,p}\hat{e}_+ + \exp(i\Delta\phi_{gp})\mathrm{LG}_{0,0}\hat{e}_-\right)$, where the intramodal phase variation depends on the topological charge of the signal, i.e., $\Delta\phi_{gp}=|\ell|\Omega/2$, and can be characterized by spatial Stokes tomography. Figure 4(b) shows the observed $\Delta\phi_{gp}$ resulting from cyclic transformation on the 4th-order modal spheres of Fig. 1(b). It is shown that for a given $\Omega$, the accumulated $\phi_{gp}$ is in direct proportion to the amount of $|\ell|$ carried by LG conjugates located on the two poles of the sphere and is independent of the radial index. This is the reason why, compared with the reference mode $\mathrm{LG}_{0,0}$, no additional phase retardation was observed on the modal sphere $\mathrm{HLG}_{2,2}$.

**Conclusion**

We have presented and experimentally demonstrated a planar zero-order astigmatic retarder fabricated with liquid-crystal geometric phase. This planar retarder was designed with the principle of fractional Fourier transformation, providing a $\lambda/4$ phase retardance between orthogonal HG modes, such that it can freely link all possible OAM states on the same HLG modal sphere via high-dimensional unitary transformation, as well as to exploit the higher-order geometric phase. Interestingly, the principle behind this planar-optics-based spatial-mode transformation is extremely similar to that of temporal-mode transformation with a time lens, it is its two-dimensional analogous in spatial degrees of freedom. Compared with previous lens-based spatial mode conversion, the geometric phase approach presented here, consists of only two identical planar optical elements, which provides good transformation performance, easy alignment, and is insensitivity to off-axis errors. This easily-deployed modal interface with planar optics provides a powerful toolkit for exploiting high-dimensional resources in OAM states and associated structured Gaussian modes.

**Methods**

**LC Geometric Phase Microstructure Fabrication:** To fabricate the LC zero-order astigmatic retarder, the nematic LC E7 (HCCH, China) was used and endowed with specific microstructures via the UV photoalignment technology. Firstly, two indium-tin oxide glasses were ultrasonic and UV-ozone cleaned. Secondly, these glasses were spin-coated with the solution of the photoalignment agent at 800 rpm for 10 s and 3000 rpm for 40 s. The used photoalignment agent solution was made up of the sulfonic azo-dye SD1 (Dai-Nippon Ink and Chemicals, Japan) dissolved in dimethylformamide at a concentration of 0.3 wt%. Thirdly, the glasses were cured at 100 °C for 10 min right after spin-coating. Fourthly, the two glasses were assembled into an empty cell using spacers, leaving a 6-μm-thick gap between the two glasses. Fifthly, the dynamic multi-step photoalignment process was carried out, using a UV photoalignment system based on a 1024 × 768 digital-micromirror-device (DMD, Discovery 3000, Texas Instruments) [28]. The pixel pitch of the DMD was originally 13.68 μm, and further reduced by a 5× objective. In the multi-step photoalignment process, each step corresponded to a certain exposure pattern and a corresponding UV linear polarization. As SD1 molecules tend to reorient perpendicular to the UV polarization direction, this system could finally imprint the desired director orientation pattern into the SD1 layers after overall exposure of multi-step patterns. Sixthly, the nematic LC E7 was filled into the empty photopatterned cell at 70 °C, and would follow the orientation of SD1 after cooling to room temperature owing to the intermolecular interactions.

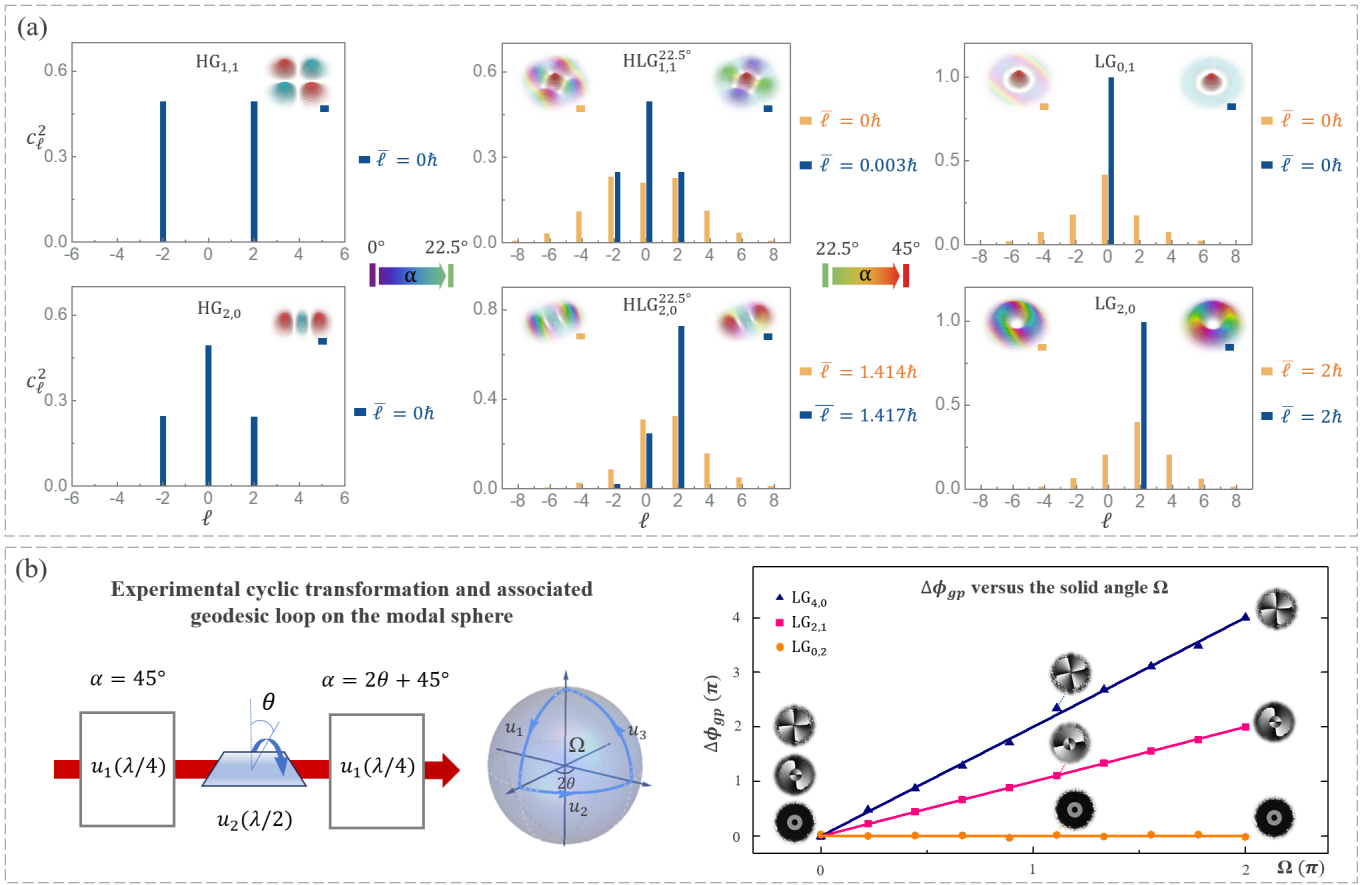

**Figure 4.** Experimental results. (a) Measured evolution of the OAM spectra in astigmatic transformation, where the spectra were calculated from the observed complex amplitude patterns embedded in the spectra. Additionally, blue and orange data correspond to input/output signals without astigmatic wavefront and astigmatic signals before passing the 2$^{nd}$ retarder, respectively; and (b) schematic and data of measured higher-order geometric phase arising from cyclic transformations on modal spheres with $N = 4$, where the geometric phase shift $\Delta\phi_{gp}$ (color dots) was obtained from the variation in the intramodal phase of vector modes, such as the example wavefronts near the data.

## Acknowledgements
This work was supported by the National Key R&D Program of China (No. 2021YFA1202000), the National Natural Science Foundation of China (Grant Nos. 62075050, 62222507, 11934013, 61975047, 12004175, and 62175101), the Innovation Program for Quantum Science and Technology (No. 2021ZD0301500), and the Natural Science Foundation of Jiangsu Province (Nos. BK20212004 and BK20200311).